\documentclass[12pt]{article}
\pdfoutput=1
\usepackage{epsfig}
\usepackage{amsfonts}
\usepackage{amssymb}
\usepackage{xcolor}

\begin{document}
\newcommand{\nc}{\newcommand}
\nc{\beq}{\begin{equation}} \nc{\eeq}{\end{equation}}
\nc{\beqa}{\begin{eqnarray}} \nc{\eeqa}{\end{eqnarray}}
\nc{\R}{{\cal R}}
\nc{\A}{{\cal A}}
\nc{\K}{{\cal K}}
\nc{\B}{{\cal B}}
\begin{center}
\begin{flushright}
Dedicated to D.V.Shirkov
\end{flushright}
\vspace{1cm}

{\bf \Large  Leading and Subleading UV Divergences in\\[0.4cm] Scattering Amplitudes for D=8 N=1 SYM Theory\\[0.4cm] in All Loops} \vspace{1.0cm}

{\bf \large D. I. Kazakov$^{1,2,3}$ and D. E. Vlasenko$^{4}$}\vspace{0.5cm}

{\it
$^1$Bogoliubov Laboratory of Theoretical Physics, Joint
Institute for Nuclear Research, Dubna, Russia.\\
$^2$Alikhanov Institute for Theoretical and Experimental Physics, Moscow, Russia\\
$^3$Moscow Institute of Physics and Technology, Dolgoprudny, Russia\\
$^4$Department of Physics, South Federal State University, Rostov-Don, Russia}
\vspace{0.5cm}

\abstract{We consider the leading and subleading UV divergences for the four-point on-shell scattering amplitudes
in D=8 N=1 supersymmetric Yang-Mills theory
in the planar limit.  This theory belongs to the class of maximally supersymmetric gauge theories  and presumably possesses distinguished properties beyond perturbation theory. We obtain the recursive relations  that allow one to get the leading and subleading divergences in all loops in a pure algebraic way staring from the one loop (for the leading poles) and two loop (for the subleading ones) diagrams. As a particular example where the recursive relations have a simple form  we consider the ladder type diagrams. The all loop summation of the leading and subleading divergences  is performed with the help of the differential equations which are the generalization of the RG equations  for non-renormalizable theories. They have explicit solutions for the ladder type diagrams.  We discuss the properties of the obtained solutions and interpretation of the results.}
\end{center}

Keywords: Amplitudes, maximal supersymmetry, UV divergences

\section{Introduction}

In the last decade there has been considerable activity on the calculation of the amplitudes in maximally supersymmetric Yang-Mills theories  (SYM)~\cite{BDS4point3loop_et_all,D=5SYM_Diverges_ZBernDixon} and maximally supersymmetric gravity~\cite{N=8SUGRA finiteness}.  Gauge and gravity SUSY theories  in $D=4$ such as the $\mathcal{N}=4$ SYM and $\mathcal{N}=8$ SUGRA are the most important examples. These theories are believed to possess several remarkable properties, among which are total or partial cancelation  of UV divergences, factorization of higher loop corrections and possible integrability. The success of factorization leading to the BDS
ansatz~\cite{BDS4point3loop_et_all} for the amplitudes in
$D=4$ $\mathcal{N}=4$ SYM stimulated similar activity in other models and dimensions.  Many magnificent insights in the structure of amplitudes (the S-matrix) of gauge theories in various dimensions (for review see, for example,~\cite{Reviews_Ampl_General}) were obtained.

In  recent papers~\cite{we1,we2} we considered the leading UV divergences of the on-shell scattering amplitudes in
maximally supersymmetric SYM theories in D=6 (N=2 SUSY), D=8 (N=1 SUSY) and D=10 (N=1 SUSY) dimensions.
In these theories the on-shell amplitudes are IR finite and the only divergences are the UV ones.  Since the gauge coupling $g^2$ in D-dimensions has dimension $[4-D]$,  all these theories are non-renormalizable.

\begin{figure}[htb]
\includegraphics[scale=0.35]{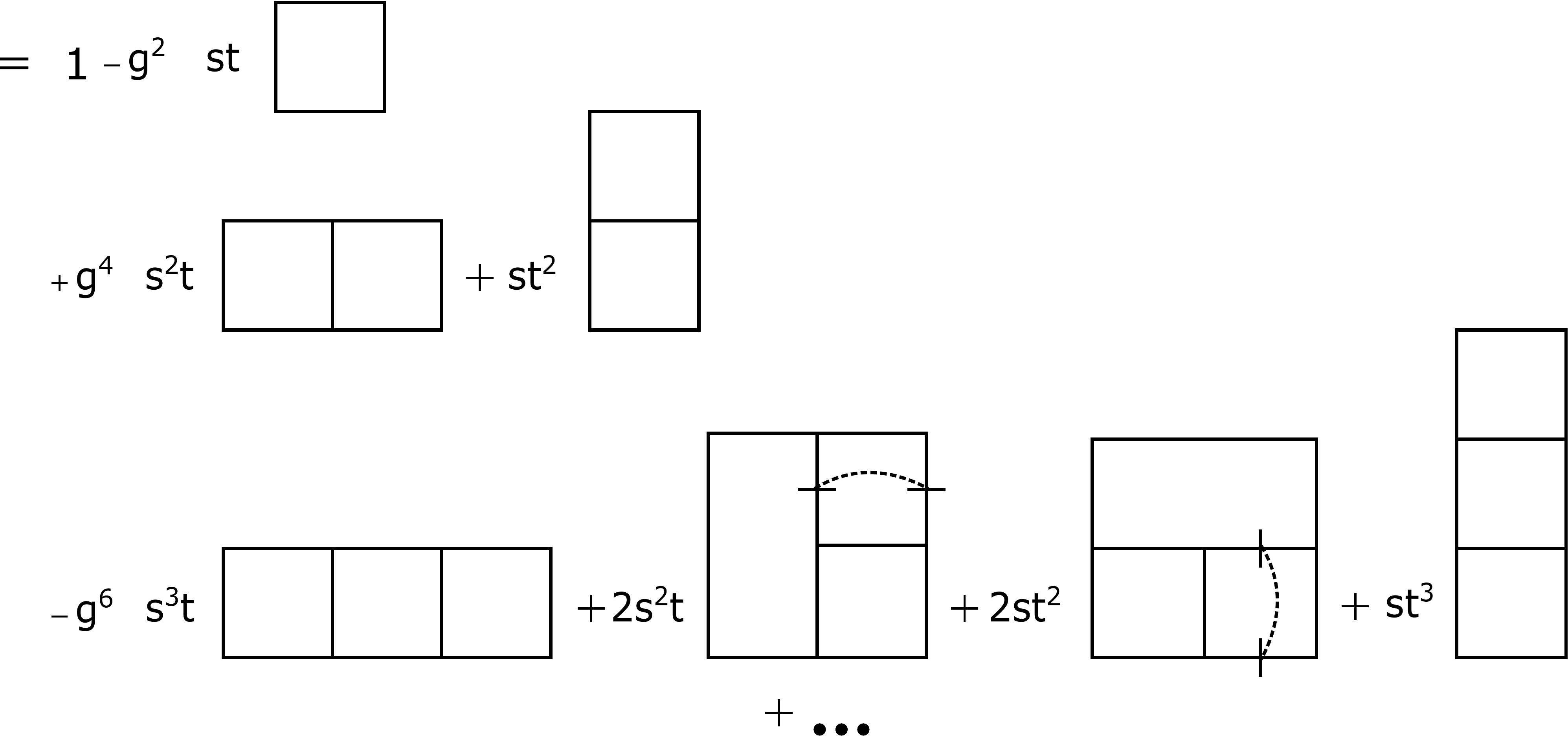} 
\caption{The universal expansion for the four-point scattering amplitude in SYM theories in terms of master integrals.
The connected strokes on the lines mean  the square of the flowing momentum.}\label{expan}
\end{figure}

Applying first the color decomposition of the amplitudes we are left with the partial amplitudes.
Within the spinor-helicity formalism~\cite{D=5SYM_Diverges_ZBernDixon} the tree level partial amplitudes have a relatively simple universal form and always factorize  so that the ratio of the loop corrections to the tree level expression can be expressed in terms of scalar master integrals.  For the four-point amplitude this is shown schematically in Fig.\ref{expan}.

\beqa
\frac{\mathcal{A}_4}{\mathcal{A}_4^{(0)}}=1+\sum\limits_L M^{(L)}_4(s,t)= \nonumber
\eeqa

The on-shell four-point amplitude depends on the Mandelstam variables s,t and u with the condition s+t+u=0.
Within the dimensional regularization (dimensional reduction) the UV divergences manifest themselves as the pole terms with the numerators being the polynomials over the kinematical variables. This expansion has a universal form in any dimension including the combinatoric factors  that  contain  the coupling constant and the powers of  s and t. The dependence on particular value of D comes from the integration inside the loops.

In D-dimensions the first UV divergences start from L=6/(D-4)  loops.
Consequently in D=6 they start from 3 loops. In D=8 and D=10, though the one loop case is  somewhat special, they start already at one loop.  In \cite{we2} we considered the leading divergences in all of these cases.
Here, we concentrate on the D=8 case since the loop order is less than in D=6 case and numerators are not so complex as in the D=10 case. Contrary to \cite{we2},  we go beyond the leading divergences and consider the subleading ones in order to understand the tendency and to check whether the
subleading terms change the situation regarding the renormalization of the theory in all loops.

\section{$\R'$-operation and pole equations  in the leading and subleading order}\label{_2}

Any local quantum field theory has a remarkable property that after performing the incomplete $\R$-operation,
the so-called $\R'$-operation, the remaining UV divergences are always local.
Let us briefly recall the main notions of the $\R$-operation~\cite{Bogolubov,Zavialov}. Being applied to any Green function  $\Gamma$ (or any particular graph $G$) it subtracts all the UV divergences including those of divergent subgraphs and leaves the finite expression. The use of the $\R$-operation is equivalent to addition of the counter terms to the initial Lagrangian. The $\R$ operation can be written in terms of subtraction operators in the factorized form
\beq  \R G = \prod_\gamma (1-M_\gamma)G, \eeq
where the subtraction operator $M_\gamma$ subtracts the UV divergence of a given subgraph $\gamma$ and the product goes over all divergent subgraphs including the graph itself.

It is useful to define also the incomplete $\R$ operation denoted by $\R'$ which subtracts only  the subdivergences of the graph $G$. The full $\R$ operation is then defined as
\begin{equation}
\R G = (1-\K) \R' G,
\end{equation}
where $\K$ is an operator that singles out the singular part of the graph (for the minimal subtraction scheme  the operator $\K$ singles out the  $1/\epsilon^n$ terms). The
$\K\R' G$  is the counter term corresponding to the  graph $G$.
Each counter term contains only the superficial divergence and is  {\it local} in coordinate space (in our case it must be a polynomial of external momenta).

The $\R'$ operation for any graph G can be defined by the forest formula, but for our calculations it is more convenient to use  the recursive definition via the $\R'$ operation for  divergent subgraphs (for details and examples see chapter 3 in \cite{Vasiliev}):
\begin{equation}
{\cal R}' G= \left(1-\sum_\gamma {\cal KR}'_\gamma +\sum_{\gamma,\gamma'}{\cal KR}'_\gamma {\cal KR}'_{\gamma'} - ...\right) G.
\label{str}
\end{equation}
The sum goes over all 1PI, UV-divergent subgraphs of the given diagram and the multiple sums include only the non-intersecting subgraphs.

When applying this formula to the diagrams at hand one finds out that for the n-loop diagram the ${\cal R}'$-operation results in the series of terms (we consider the leading and subleading poles)
\beqa
{\cal R'}G_n&=&\frac{\A_n^{(n)}(\mu^2)^{n\epsilon}}{\epsilon^n}+\frac{\A_{n-1}^{(n)}(\mu^2)^{(n-1)\epsilon}}{\epsilon^n}+ ... +\frac{\A_1^{(n)}(\mu^2)^{\epsilon}}{\epsilon^n}\nonumber \\
&+&\frac{\B_n^{(n)}(\mu^2)^{n\epsilon}}{\epsilon^{n-1}}+\frac{\B_{n-1}^{(n)}(\mu^2)^{(n-1)\epsilon}}{\epsilon^{n-1}}+ ... +\frac{\B_1^{(n)}(\mu^2)^{\epsilon}}{\epsilon^{n-1}} \nonumber \\
&+&\mbox{lower\ order\ terms,}\label{Rn}
\eeqa
where the terms like $\frac{\A_{k}^{(n)}(\mu^2)^{k\epsilon}}{\epsilon^n}$  and $\frac{\B_{k}^{(n)}(\mu^2)^{k\epsilon}}{\epsilon^{n-1}}$ come from the $k$-loop graph which survives after subtraction of the $(n-k)$-loop counterterm.
 The full expression (\ref{Rn}) has to be local, i.e. should not contain terms like $(\log \mu)^k/\epsilon^m$ for all $k,m>0$ while being expanded over $\epsilon$. (For simplicity hereafter we put $\mu^2 \equiv \mu$.)
This requirement gives us $n-1$ equations for the coefficients $\A_i^{(n)}$ and $n-2$ equations for the coefficients $\B_i^{(n)}$. Solving them in favour of the one and two loop graphs one gets
\beqa
\A_n^{(n)}&=&(-1)^{n+1}\frac{\A_1^{(n)}}{n}, \label{rel}\\
\B_n^{(n)}&=&(-1)^n \left(\frac 2n \B_2^{(n)}+\frac{n-2}{n}\B_1^{(n)}\right) \label{rel1}.
\eeqa
It is also useful to have analogous expressions for the ${\cal KR'}$ terms equal to
\beq
{\cal KR'}G_n=\sum_{k=1}^n \left(\frac{\A_k^{(n)}}{\epsilon^n} +\frac{\B_k^{(n)}}{\epsilon^{n-1}}\right)\equiv
\frac{\A_n^{(n)'}}{\epsilon^n}+\frac{\B_n^{(n)'}}{\epsilon^{n-1}}.
\eeq
One has, respectively,
\beqa
\A_n^{(n)'}&=&(-1)^{n+1}\A_n^{(n)}=\frac{\A_1^{(n)}}{n}, \label{rel2} \\
\B_n^{(n)'}&=& \left(\frac{2}{n(n-1)} \B_2^{(n)}+\frac{2}{n}\B_1^{(n)}\right) \label{rel3}.
\eeqa
This means that performing the  ${\cal R}'$-operation one can take care only of the one loop  diagrams
surviving after contraction and get the desired leading pole term via eq.(\ref{rel}) and add the two loop diagrams to get the subleading pole from eq.(\ref{rel1}). This observation drastically simplifies the calculation of the leading and subleading poles. Moreover, it follows from eqs.(\ref{rel},\ref{rel1}) that, just as in renormalizable theories, the  leading poles are essentially governed by the one loop diagrams and can be deduced from them  in all loops pure algebraically, while to know the subleading poles in all loops one needs to know the subleading pole of the two loop diagrams. We demonstrate below how this procedure works in practice and obtain explicit formulas for the leading and subleading poles in all loops.

\section{The leading poles in all loops}
\begin{figure}[ht!]
\begin{center}
\leavevmode
\includegraphics[width=0.8\textwidth]{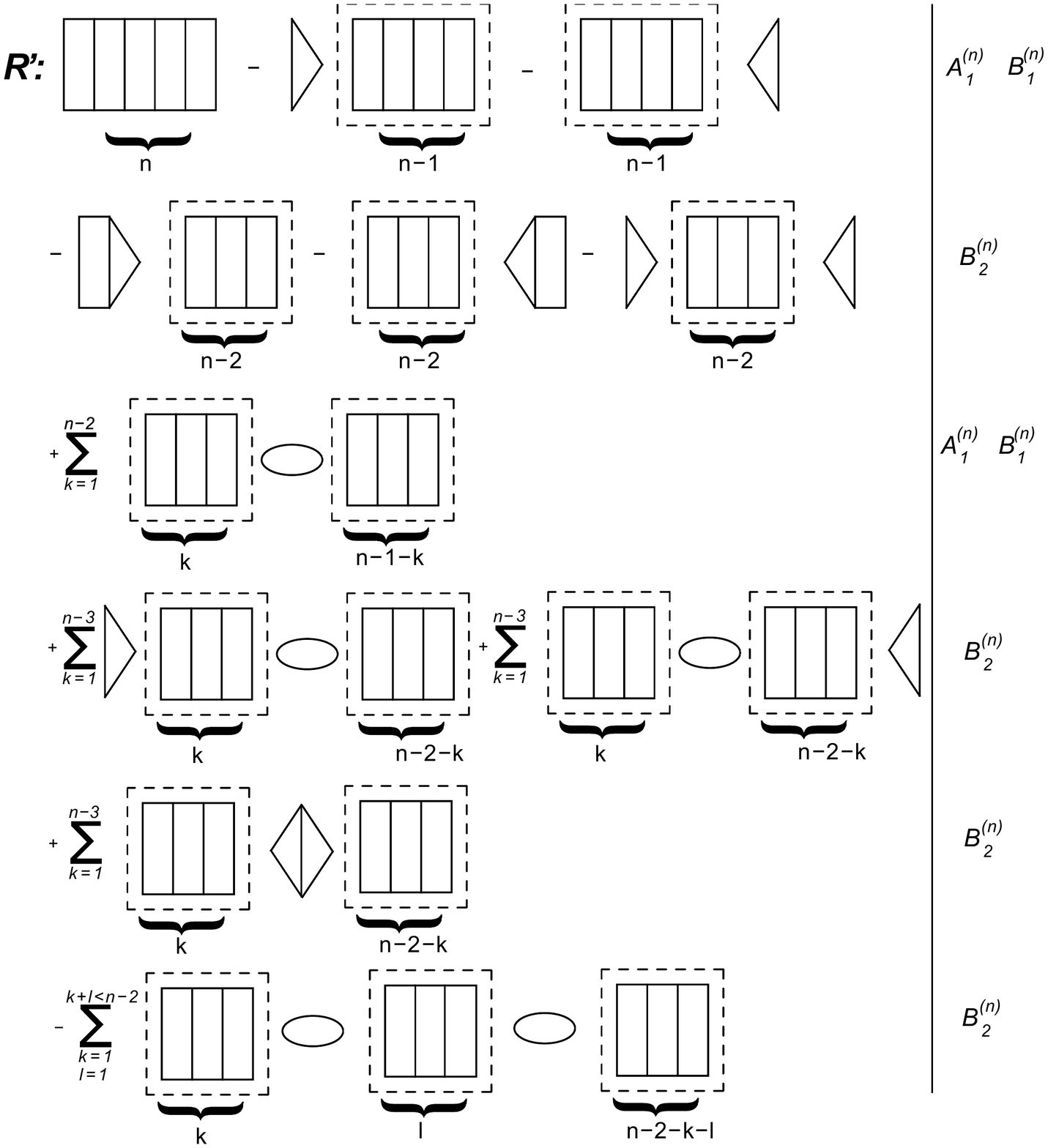}
\end{center}\vspace{-3.5cm}
\caption{The application of the $\R'$-operation to the ladder type diagrams. At the right it is shown to which term
of expansion (\ref{Rn}) it corresponds to}
\label{ladder}
\end{figure}

We start with the leading poles and consider as a simplest example the $\R'$-operation applied to the ladder type diagrams (see Fig.\ref{ladder}).

To calculate the contribution to $\A_1^{(n)}$ one has to calculate the poles of the
one-loop  diagrams in the first and the third row.  For $\B_1^{(n)}$ one needs the constant part of these one loop graphs
and for $\B_2^{(n)}$ one needs the leading and subleading poles of the two-loop diagrams from the second and  fourth  row.

We first concentrate on the leading poles. For the s-channel ladder type diagram they depend only on $s$ so that $\A_n^{(n)}=s^{n-1} A_n$. Calculating the needed one loop diagrams and substituting them into eq.(\ref{rel}) one gets the recursion relation
\beq
n A_n=-\frac{2}{4!} A_{n-1}+\frac{2}{5!}\sum_{k=1}^{n-2}A_kA_{n-1-k}, \ \ \  n\geq 3 \label{one8}
\eeq
with $A_1=1/3!$. Starting from this value one can calculate any $A_n$ though explicit solution of the recursion relation (\ref{one8}) is not straightforward.

 However, since we actually need the sum of the series we perform the summation multiplying both sides of eq.(\ref{one8}) by $(-z)^{n-1}$, where $z=\frac{g^2}{\epsilon}$ and take the sum  from 3 to infinity
\beq
\sum_{n=3}^\infty  n A_n (-z)^{n-1}=-\frac{2}{4!}\sum_{n=3}^\infty A_{n-1} (-z)^{n-1}+\frac{2}{5!}\sum_{n=3}^\infty
\sum_{k=1}^{n-2}A_k (-z)^k A_{n-1-k} (-z)^{n-1-k}. \label{sum}
\eeq
Denoting now the sum $\sum_{n=m}^\infty A_n (-z)^n$ by $\Sigma_m$ and performing the interchange of the order of summation in the nonlinear term we get
\beq
-\frac{d}{dz}\Sigma_3=-\frac{2}{4!}\Sigma_2+\frac{2}{5!}\Sigma_1\Sigma_1.
\eeq

Having in mind that
$$\Sigma_3=\Sigma_1+A_1 z-A_2 z^2, \ \ \Sigma_2=\Sigma_1+A_1 z, \ \ A_1=\frac{1}{3!},\ A_2=-\frac{1}{3!4!},$$
one finally gets the equation for $\Sigma_A\equiv\Sigma_1$
\beq
\frac{d}{dz}\Sigma_A=-\frac{1}{3!}+\frac{2}{4!}\Sigma_A-\frac{2}{5!}\Sigma_A^2. \label{eqa}
\eeq
Solution to this equation is
\beq
\Sigma_A(z)=-\sqrt{5/3} \frac{4 \tan(z/(8 \sqrt{15}))}{1 - \tan(z/(8 \sqrt{15})) \sqrt{5/3}}=\sqrt{10}\frac{\sin(z/(8\sqrt{15}))},{\sin(z/(8\sqrt{15})-z_0)}\label{sol},
\eeq
where $z_0=\arcsin(\sqrt{3/8})$.
The expansion of $\tan z$ contains the Bernoulli numbers
$$ \tan z = \sum_{n=1}^\infty (-1)^{n-1}\frac{2^{2n}(2^{2n}-1)B_{2n}}{(2n)!}z^{2n-1}.$$
Being substituted into eq.(\ref{sol}) it gives
\beq \Sigma_A(z)=-(z/6+z^2/144+z^3/2880+7z^4/414720 + \dots) \label{expansion}. \eeq
Note that taking into account the combinatoric factors the dimensionless variable  $z$ is given by $z = \frac{g^2 s^2}{\epsilon}$.
This series reproduces the ladder type diagrams in all orders. The same is obviously true for the t-channel ladder  diagram with the replacement $s\leftrightarrow t$.
The function $\Sigma_A$  given by eq.(\ref{sol})  has an infinite sequence of simple poles and thus has no limit when $z\to\infty$ i.e., $\epsilon \to 0$.

 The recursion relation (\ref{one8}) can be generalized  to include all diagrams contributing to the four point amplitude.  It
 was obtained in \cite{we2} and we present here a short summary.
 It is worth mentioning that not all the 4-point diagrams contribute to the leading pole but only those which contain the subgraphs of each previous order. These diagrams can be constructed with the help of the so called rung-rule, described in \cite{we2}. But even among them there are zero contributions which come from the diagrams that are reduced to the two-point functions after shrinking the subgraphs when performing the $\R'$-operation.

To present the recursion relation for the full set of relevant diagrams we  note that they  can be divided into two classes: the s-channel and the t-channel ones and the total contribution is the sum of them. The singularity  of $G_{n}$ is a polynomial in $s$ and $t$. When calculating $\A_1^{(n)}$  like in Fig.\ref{ladder} for the s-channel diagrams, $s$ stands as  a constant factor while $t$ from ${\cal K R}'G_{n-1}$ contains the integration  momentum over the last loop. The same is true for  the t-channel diagrams  with the replacement $s\leftrightarrow t$.

 As a result, when using the
 $\R'$-operation one is left  with the remaining triangle and bubble one loop diagrams shown in the first and the third rows of Fig.\ref{ladder}.  Substituting the explicit form of $s$ and $t$ and integrating over the triangle and the bubble with the help of the  Feynman parameters one gets the desired recursion relation.
 Note that for the bubble term when integrating over the loop on both sides one has functions of $s$ and $t$. Replacing $t$ by $t'$ one should have in mind that on the left $t'=(l-p_1)^2$ and on the right $t'=(l+p_4^2)$, where $l$ is the integration momentum. This means that while integration one gets the mixed terms like $g^{\mu\nu}p_1^\mu p_4^\nu$.
 They give rise to the double sum in the second term of the equation.
 Eventually one has
\beqa
&&nS_n(s,t)=-2 s^2 \int_0^1 dx \int_0^x dy\  y(1-x) \ (S_{n-1}(s,t')+T_{n-1}(s,t'))|_{t'=tx+yu}\label{eq8}\nonumber \\ &+&
s^4 \int_0^1\! dx \ x^2(1-x)^2 \sum_{k=1}^{n-2}  \sum_{p=0}^{2k-2} \frac{1}{p!(p+2)!} \
 \frac{d^p}{dt'^p}(S_{k}(s,t')+T_{k}(s,t')) \times \nonumber \\
&&\hspace{2cm}\times  \frac{d^p}{dt'^p}(S_{n-1-k}(s,t')+T_{n-1-k}(s,t'))|_{t'=-sx} \ (tsx(1-x))^p,
\eeqa
where $t'= t x+u y$, $u=-t-s$, and $S_1= \frac{1}{12},\ T_1=\frac{1}{12}$.

Here we denote by  $S_n(s,t)$ and  $T_n(s,t)$  the sum of all contributions  in the  $n$-th order of PT in $s$ and  $t$ channels, respectively.

The same recursive relation is valid for the $t$-channel  diagrams with the obvious replacement $s\leftrightarrow t$. Due to  the $s-t$ symmetry of the amplitude, one should have $T_n(s,t)=S_n(t,s)$.
The coefficient $\A_n^{(n)}(s,t)$ of the $n$'th order pole is the sum
$$
\A_n^{(n)}(s,t)=S_n(s,t)+T_n(s,t).
$$

Equation (\ref{eq8}) can be summed the same way as in the ladder case (\ref{sum}). Multiplying both sides by  $(-z)^{n-1}$ and summing up over $n$ from 3 to  infinity one gets
\beqa
&&\frac{d}{dz}\Sigma_3(s,t,z)=2 s^2 \int_0^1 dx \int_0^x dy\  y(1-x)\ (\Sigma_2(s,t',z)+\Sigma_2(t',s,z)
|_{t'=tx+yu}\\
&&-s^4  \int_0^1\! dx \ x^2(1-x)^2 \sum_{p=0}^\infty \frac{1}{p!(p+2)!} (\frac{d^p}{dt'^p}(\Sigma_1(s,t',z)+\Sigma_1(t',s,z)|_{t'=-sx})^2 \ (tsx(1-x))^p. \nonumber
\eeqa
Using now that
$$ \Sigma_3(s,t,z)=\Sigma_1(s,t,z)-S_2(s,t) z^2+S_1(s,t)z, \  \Sigma_2(s,t,z)=\Sigma_1(s,t,z) +S_1(s,t)z, $$
$$\frac{d}{dz}\Sigma_3(s,t,z)= \frac{d}{dz}\Sigma_1(s,t,z)- 2S _ 2(s,t)z+S_1(s,t), \ \  2S_2(s,t)=2s^2\int (S_1(s,t')+S_1(t',s))$$
and making notation $\Sigma(s,t,z)=\Sigma_1(s,t,z)$ one finally obtains
\beqa
&&\frac{d}{dz}\Sigma(s,t,z)=-\frac{1}{12}+2 s^2 \int_0^1 dx \int_0^x dy\  y(1-x)\ (\Sigma(s,t',z)+\Sigma(t',s,z))|_{t'=tx+yu}
\label{tot}\\
&&-s^4  \int_0^1\! dx \ x^2(1-x)^2 \sum_{p=0}^\infty \frac{1}{p!(p+2)!} (\frac{d^p}{dt'^p}(\Sigma(s,t',z)+\Sigma(t',s,z))|_{t'=-sx})^2 \ (tsx(1-x))^p. \nonumber
\eeqa
The same equation with the replacement $s \leftrightarrow t$ can be derived for $\Sigma(t,s,z)$.

Contrary to the ladder case (\ref{eqa}), solution of  equation  (\ref{tot}) is not straightforward and difficult to analyze.

\section{The subleading poles in all loops. The ladder diagrams}

We now turn to the subleading pole. As an example we again consider the ladder diagrams. Contrary to the leading case,
here the input is defined not only  by the one loop diagrams, but by the two-loop ones as well. The subleading pole of the two loop box has to be explicitly evaluated, it cannot be deduced from the recursion  relations as it follows from eq.
(\ref{rel1}). Explicit evaluation gives
\beq
Box=\frac{1}{3!\epsilon}, \ \ \  \ DoubleBox=-\frac{1}{3!4!}(\frac{s}{\epsilon^2}+\frac{27}{4}\frac{s}{\epsilon}+\frac 16\frac{t}{\epsilon}).
\eeq
Together with the other one and two loop diagrams evaluated up to the subleading order we obtain the expressions for $\B_1^{(n)}$ and $\B_2 ^{(n)}$ from the diagrams of Fig.\ref{ladder}. The evaluation of some two loop diagrams is quite a cumbersome task;  for instance, the rhombus one has been calculated using the integration by parts technique~ \cite{calc}. Hereafter we use the notation
$$\A_n^{(n)}=s^{n-1}A_n, \   \A_n^{(n)'}=s^{n-1}A'_n,\ \B_n^{(n)}=s^{n-1}B_{sn}+s^{n-2}tB_{tn}, \ \B_n^{(n)'}=s^{n-1}B'_{sn}+s^{n-2}tB'_{tn}.
$$
We get
\beqa
\B_1^{(n)}&=&-A'_{n-1}s^{n-2}(-\frac{s}{4!})\frac{19}{6}2-B'_{sn-1}s^{n-2}(-\frac{s}{4!})2-B'_{tn-1}s^{n-3}(-\frac{s}{5!})(t-2s)2\nonumber \\
&+& \sum_{k=1}^{n-2}A'_ks^{k-1}A'_{n-1-k}s^{n-2-k}(\frac{2s^2}{5!})\frac{46}{15}\\
&+& \sum_{k=1}^{n-2}A'_ks^{k-1}B'_{sn-1-k}s^{n-2-k}(\frac{2s^2}{5!})2
+\sum_{k=1}^{n-2}A'_ks^{k-1}B'_{tn-1-k}s^{n-3-k}(\frac{-s^3}{5!})2, \nonumber
\eeqa
\beqa
\B_2^{(n)}&=&-A'_{n-2}s^{n-3}(\frac{s^2}{3!4!})\frac{5063}{2400}2-B'_{sn-2}s^{n-3}(\frac{s^2}{3!4!})\frac{13}{40}2-B'_{tn-1}s^{n-4}(\frac{s^2}{5!5!})\frac{t-32s}{2}2\nonumber \\
&-&A'_{n-2}s^{n-3}(-\frac{s}{4!})(-\frac{s}{4!}\frac{19}{6})2-B'_{sn-2}s^{n-3}(-\frac{s}{4!})(-\frac{s}{4!})-B'_{tn-2}s^{n-4}(-\frac{s^2}{5!5!})(12s-t)\nonumber\\
&+& \sum_{k=1}^{n-3}A'_ks^{k-1}A'_{n-2-k}s^{n-3-k}(-\frac{s^3}{5!4!})\frac{938}{15}\nonumber\\
&+& \sum_{k=1}^{n-3}A'_ks^{k-1}B'_{sn-2-k}s^{n-3-k}(-\frac{s^3}{5!2})
+\sum_{k=1}^{n-3}A'_ks^{k-1}B'_{tn-2-k}s^{n-4-k}(\frac{442 s^4}{5!5!12}) \nonumber\\
&-&\sum_{k,l=1}^{n-k+l<n-2}A'_ks^{k-1}A'_ls^{l-1}A'_{n-2-k-l}s^{n-3-k-l}(\frac{2s^2}{5!})(\frac{2s^2}{5!})\frac{46}{15}2
\nonumber\\
&-&\sum_{k,l=1}^{n-k+l<n-2}A'_ks^{k-1}A'_ls^{l-1}B'_{sn-2-k-l}s^{n-3-k-l}(\frac{2s^2}{5!})(\frac{2s^2}{5!})3\nonumber  \\
&-&\sum_{k,l=1}^{n-k+l<n-2}A'_ks^{k-1}A'_ls^{l-1}B'_{tn-2-k-l}s^{n-4-k-l}(\frac{2s^2}{5!})(-\frac{s^3}{5!})2 \nonumber \\
&-&\sum_{k,l=1}^{n-k+l<n-2}A'_ks^{k-1}B'_{tl}s^{l-2}A'_{n-2-k-l}s^{n-3-k-l}(\frac{-2s^5}{5!5!}).
\eeqa
These expressions it their turn allow us to get the desired recursion relations for the primed coefficients using eqs.(\ref{rel3})
\beqa
B'_{tn}&=&-\frac{2}{n(n-1)}B'_{tn-2}\frac{10}{5!5!}+\frac 2n B'_{tn-1}\frac{2}{5!}, \label{btprime} \\
B'_{sn}&=&\frac{2}{n(n-1)}\left[-A'_{n-2}\frac{2321}{5!5!2}-B'_{sn-2}\frac{18}{4!5!}+B'_{tn-2}\frac{44}{5!5!}\right. \nonumber\\
&-& \left. \sum_{k=1}^{n-3}A'_kA'_{n-2-k}\frac{938}{4!5!15}- \sum_{k=1}^{n-3}A'_kB'_{sn-2-k}\frac{1}{5!2}+
 \sum_{k=1}^{n-3}A'_kB'_{tn-2-k}\frac{442}{5!5!12}\right. \nonumber\\
 &-&\left. \sum_{k,l=1}^{n-k+l<n-2}A'_kA'_lA'_{n-2-k-l}\frac{8}{5!5!}\frac{46}{15}
  -\sum_{k,l=1}^{n-k+l<n-2}A'_kA'_lB'_{sn-2-k-l}\frac{12}{5!5!}\right. \nonumber \\
 &+&\left. \sum_{k,l=1}^{n-k+l<n-2}A'_kA'_lB'_{tn-2-k-l}\frac{4}{5!5!}
  +\sum_{k,l=1}^{n-k+l<n-2}B'_kA'_lA'_{sn-2-k-l}\frac{2}{5!5!}\right]\nonumber \\
  &+&\frac 2n\left[ A'_{n-1}\frac{19}{3 4!}+B'_{sn-1}\frac{2}{4!}-B'_{tn-1}\frac{4}{5!}\right. \nonumber \\
 &+& \left. \sum_{k=1}^{n-2}A'_kA'_{n-1-k}\frac{2}{5!}\frac{46}{15}+\sum_{k=1}^{n-2}A'_kB'_{sn-1-k}\frac{4}{5!}-
 \sum_{k=1}^{n-2}A'_kB'_{tn-1-k}\frac{2}{5!}\right] .\label{bsprime}
\eeqa
One can write down the corresponding relations for the unprimed coefficients in terms of the primed ones. For $B_{tn}$
it looks like
\beq
B_{tn}=(-1)^n\left[ -\frac{2}{n}B'_{tn-2}\frac{10}{5!5!}+\frac{n-2}{n} B'_{tn-1}\frac{2}{5!}\right] \label{bt}
\eeq
and similar for $B_{sn}$. Starting from the initial values
$$B'_{s1}=B'_{t1}=0, \ \  B'_{s2}=-\frac{1}{3!4!}\frac{5}{12}, \ B'_{t2}=-\frac{1}{3!4!}\frac{1}{6},  B_{s2}=-\frac{1}{3!4!}\frac{27}{4},\ B_{t2}=-\frac{1}{3!4!}\frac{1}{6},\ A_1=\frac{1}{3!}$$
they allow one to calculate the corresponding coefficients at any order. One can write down a simple Mathematica routine to evaluate them pure algebraically (see attachment Mathematica file in Supplementary Material)

The last step is to sum the series. In order to do it we again write down the differential equation for the sum.
Let us begin with $B'_{sn}$. Multiplying eq.(\ref{btprime}) by $z^{n-2}$  and taking the sum from 3 to infinity
we get the differential equation for $\Sigma'_{tB}\equiv \ \Sigma'_{tB2}=\sum_2^\infty z^nB'_{tn}$
\beq
\frac{d^2 \Sigma'_{tB}(z)}{dz^2}-\frac{1}{30}\frac{d \Sigma'_{tB}(z)}{dz}+\frac{\Sigma'_{tB}(z)}{720}=-\frac{1}{432}
\label{eq1}
\eeq
in full analogy with eq.(\ref{eqa}). Having in mind the boundary conditions $\Sigma'_{tB}(0)=\frac{d \Sigma'_{tB}(0)}{dz}=0$, one gets the solution
\beq
\Sigma'_{tB}(z)=\frac{5}{6}\left[e^{z/60}(-sin[z/30]+2cos[z/30])-2\right].\label{solp}
\eeq
One can write down the equation for the unprimed $\Sigma_{tB}\equiv \ \Sigma_{tB2}=\sum_2^\infty (-z)^nB_{tn}$. It follows from eq.(\ref{bt}) and has the form
  \beq
\frac{d\Sigma_{tB}(z)}{dz}=\frac{1}{60}z\frac{d \Sigma'_{tB}(z)}{dz}-\frac{\Sigma'_{tB}(z)}{60}-z\frac{\Sigma'_{tB}(z)}{720}-\frac{z}{432}.\label{eq}
\eeq
Solving this equation having in mind eq.(\ref{solp}) one gets
\beq
\Sigma_{tB}(z)=-\frac{1}{36}\left[60 + z + e^{z/60} (-(60 + z) cos[z/30] - 2 (-15 + z) sin[z/30])\right] .\label{sol2}
\eeq

It is more difficult to get the closed expressions for $\Sigma'_{sB}$ and  $\Sigma_{sB}$. Again we start with the primed coefficients. Taking eq.(\ref{bsprime}) as input, multiplying it by $z^{n-2}$ and taking the sum from 3 to infinity, we get the following differential equation for $\Sigma'_{sB}\equiv \ \Sigma'_{sB2}=\sum_2^\infty z^nB'_{sn}$
\beq
\frac{d^2 \Sigma'_{sB}(z)}{dz^2}+f_1(z)\frac{d \Sigma'_{sB}(z)}{dz}+f_2(z)\Sigma'_{sB}(z)=f_3(z), \label{Ric}
\eeq
where
\beqa
f_1(z)&=&-\frac 16+\frac{\Sigma_A}{15},\nonumber\\
f_2(z)&=&\frac{1}{80}-\frac{\Sigma_A}{120}+\frac{\Sigma_A^2}{600}+\frac{1}{15}\frac{d \Sigma_A}{dz}, \nonumber\\
f_3(z)&=&\frac{2321}{5!5!2}\Sigma_A+\frac{11}{1800}\Sigma'_{tB}-\frac{469}{5!90}\Sigma_A^2-\frac{442}{5!5!6}\Sigma_A\Sigma'_{tB}+\frac{23}{6750}\Sigma_A^3+\frac{1}{1200}\Sigma_A^2\Sigma'_{tB}\nonumber\\
&-&\frac{19}{36}\frac{d \Sigma_A}{dz}-\frac{1}{15}\frac{d \Sigma'_{tB}}{dz}
+\frac{23}{225}\frac{d \Sigma_A^2}{dz}+\frac{1}{30}\frac{d( \Sigma_A\Sigma'_{tB})}{dz}-\frac{3}{32}.\nonumber
\eeqa
This is a general Riccati equation. Surprisingly the solution can be found by a simple substitution. We discuss it in the next section.

Similarly to eq.(\ref{eq}) one can write down the equation for $\Sigma_{sB}\equiv \ \Sigma_{sB2}=\sum_2^\infty (-z)^nB_{sn}$.  It can be solved in terms of $\Sigma'_{sB}$ and $\Sigma'_{tB}$ in a straightforward way
\beq
\Sigma_{sB}=(z\frac{d}{dz}-1)\Sigma'_{sB}-z(-\frac{19}{72}\Sigma_A+\frac{1}{12}\Sigma'_{sB}-\frac{1}{30}\Sigma'_{tB}+\frac{23}{450}\Sigma_A^2-\frac{1}{30}\Sigma_A \Sigma'_{sB}+\frac{1}{60}\Sigma_A \Sigma'_{tB}) .\label{Ric2}
\eeq

\section{Properties of the Solutions}

\begin{figure}[ht!]
\begin{center}
\leavevmode
\includegraphics[width=0.5\textwidth]{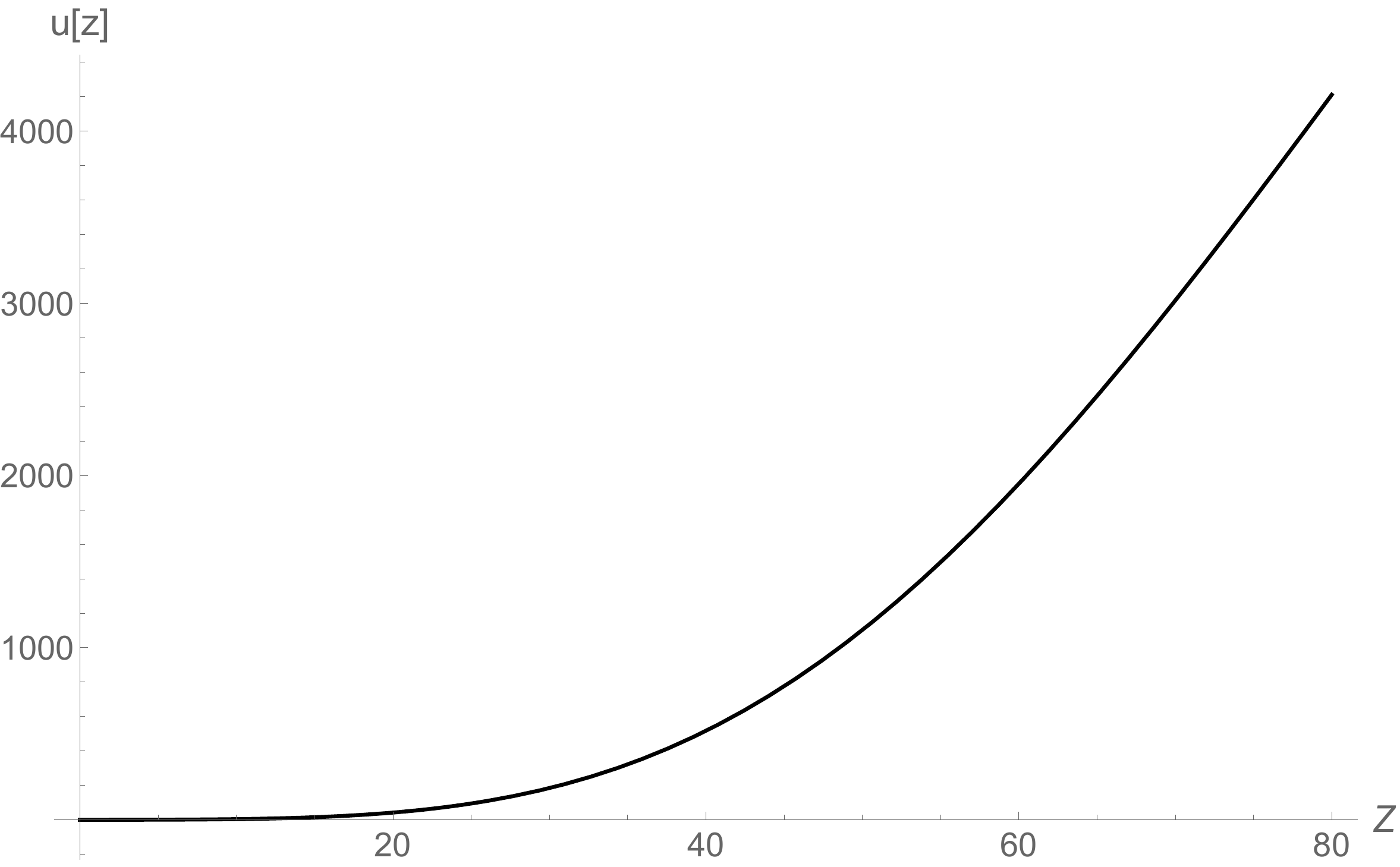}\hspace{1cm}
\end{center}
\caption{ The behaviour of u(z) evaluated numerically}
\label{uplot}
\end{figure}

We discuss here some properties of solutions of eqs.(\ref{Ric},\ref{Ric2}).
Equation (\ref{Ric}) is a linear inhomogeneous second order differential equation. Surprisingly enough it and can be simplified making the substitution: $\Sigma'_{sB}(z)=\frac{d\Sigma_A}{dz} u(z)$. Then, using equation (\ref{eqa}), it is reduced to a simple one for the function $u(z)$:
\beq
u''(z)\frac{d\Sigma_A}{dz}=f_3(z),\label{ddu}
\eeq
which is trivially solvable in quadratures
\beq
u(z)=\int_{0}^{z}dy\int_{0}^{y}dx\frac{f_3(x)}{\Sigma_A'(x)}.\label{u}
\eeq
The numerical integration demonstrates that u(z) is smooth function (see Fig.\ref{uplot}) which increases $~Exp[z/60]$ just like $\Sigma'_{tB}$ and has no poles  so that the singularities of $\Sigma'_{sB}(z)$ come only from $\Sigma'_A$.

To get the function $\Sigma_{sB}$ one has to substitute the functions $\Sigma'_{sB}$ and $\Sigma'_{tB}$ into eq.(\ref{Ric2}). One can do it numerically. The singularities of the obtained function are governed by the leading term
$\Sigma_A$ and its derivative. Consider them in more detail.

\begin{figure}[ht!]
\begin{center}
\leavevmode
\includegraphics[width=0.45\textwidth]{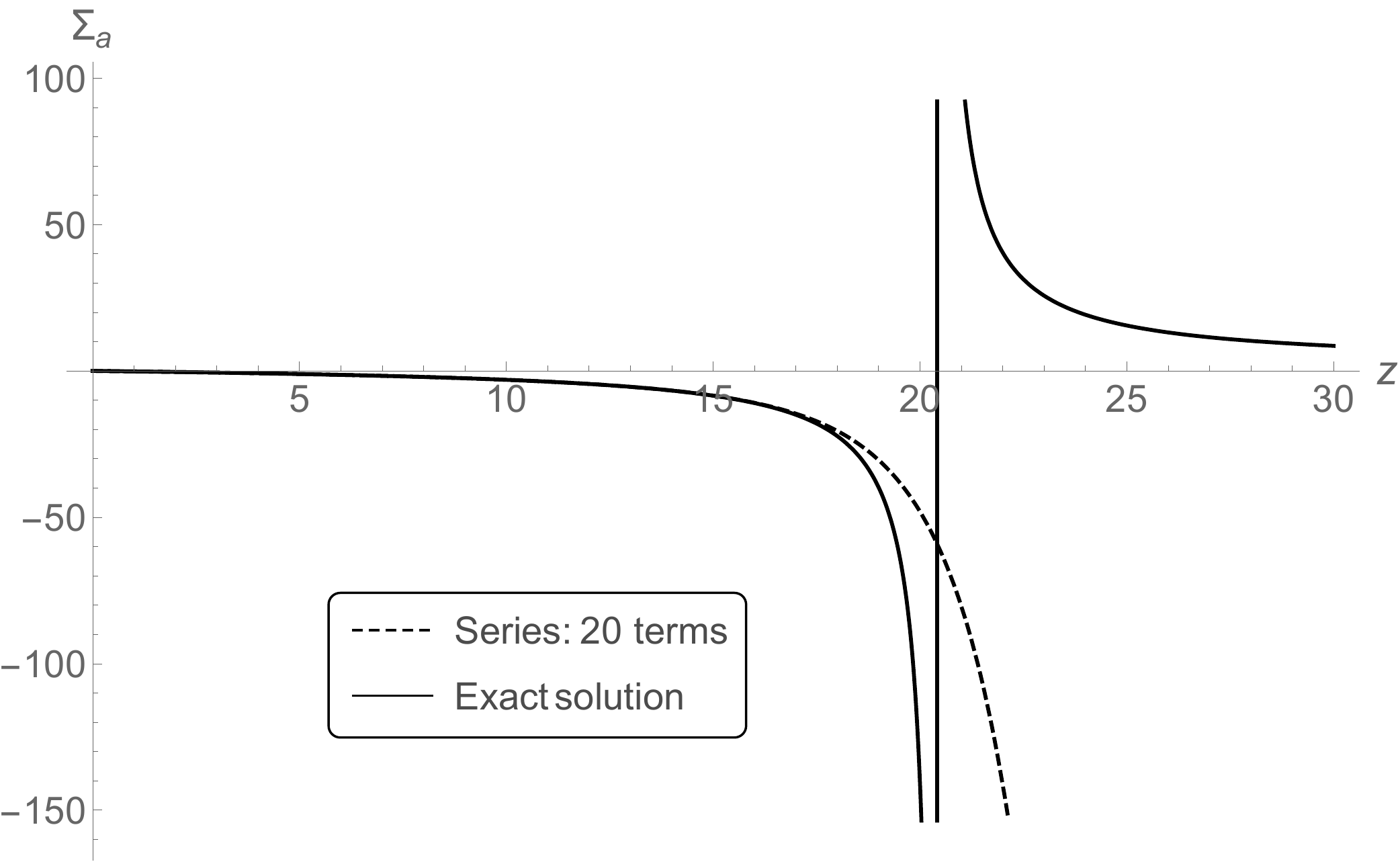}\hspace{1cm}
\includegraphics[width=0.45\textwidth]{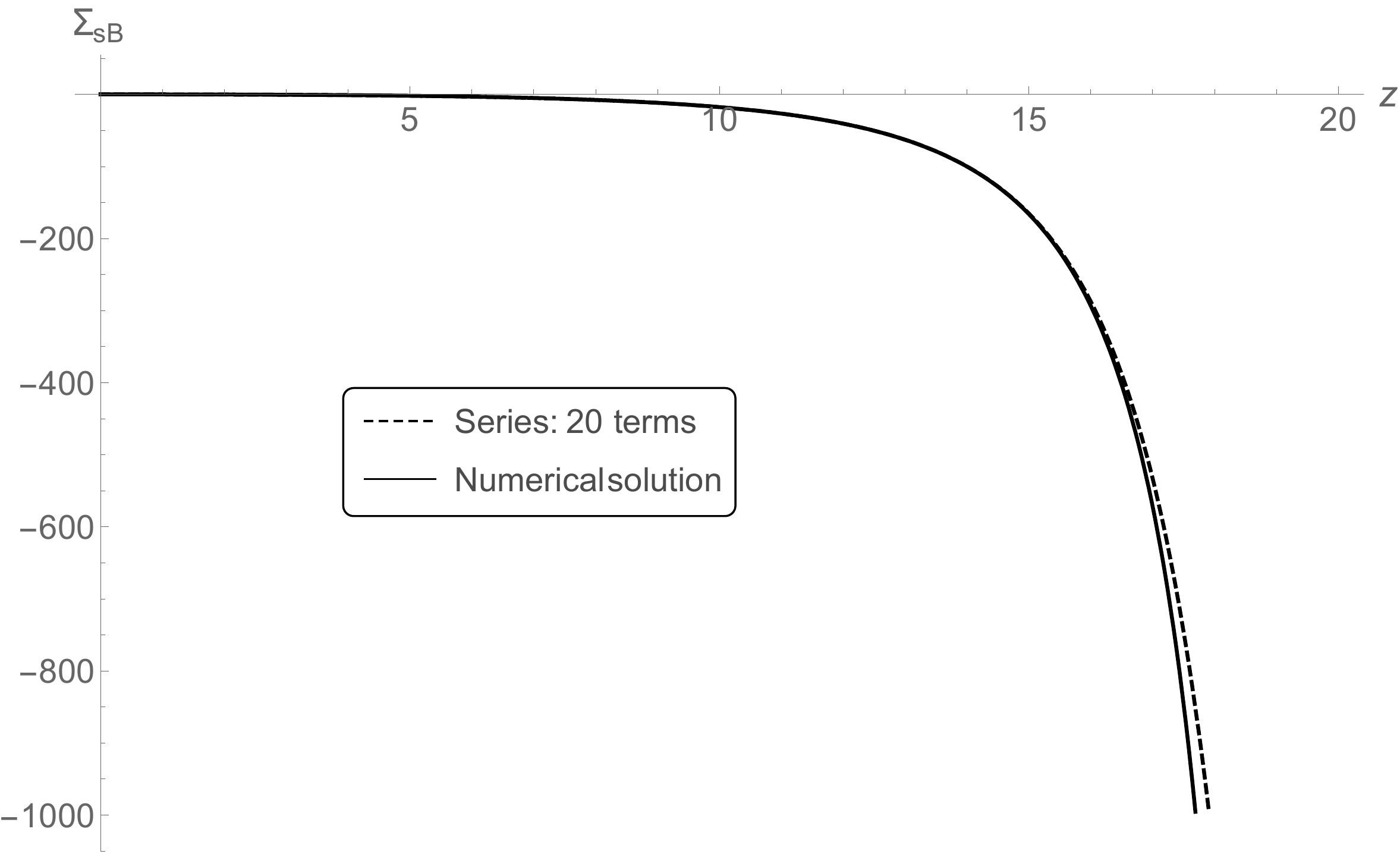}
\end{center}
\caption{Comparison of the exact solution with the perturbation series including 20 terms  for $\Sigma_A$ (left) and for
 $\Sigma_{sB}$ (right). The solid line is the exact (numerical) solution and the  dotted line is the PT series summation result}
\label{Sigma_a}
\end{figure}

The leading divergences are given by  $\Sigma_A$ (\ref{sol}).  It is a singular periodic function which has
zeroes at $z/8/\sqrt{15}=\pi n$ and first order poles at $z/8/\sqrt{15}=\arcsin\sqrt{3/8} +\pi n$. Summation of perturbative expansion (\ref{expansion}) gives a satisfactory approximation between zero  and the  first pole at $z\approx 20.42$, however fails to go beyond as can be seen in Fig.\ref{Sigma_a} (left)  below.

Turning to subleading divergences given by $\Sigma_{sB}$ (\ref{Ric2}) we notice that eq.(\ref{Ric2}) contains the singular functions $\Sigma_A$ and $\Sigma'_{sB}$ which possess the poles at the same points. Thus, we expect that the function $\Sigma_{sB}$ is a singular function of the same kind and also contains poles. Hence,
the numerical solution is due to be valid in the interval between zero and the first pole.
In the same interval one can also perform a summation of perturbation series generated with the help of the code mentioned above. We present in Fig.\ref{Sigma_a} (right) the result of  numerical solution of eqs.(\ref{Ric},\ref{Ric2})   together with the perturbation theory series taking into account 20 terms of expansion.

\begin{figure}[htb]
\begin{center}
\leavevmode
\includegraphics[width=0.6\textwidth]{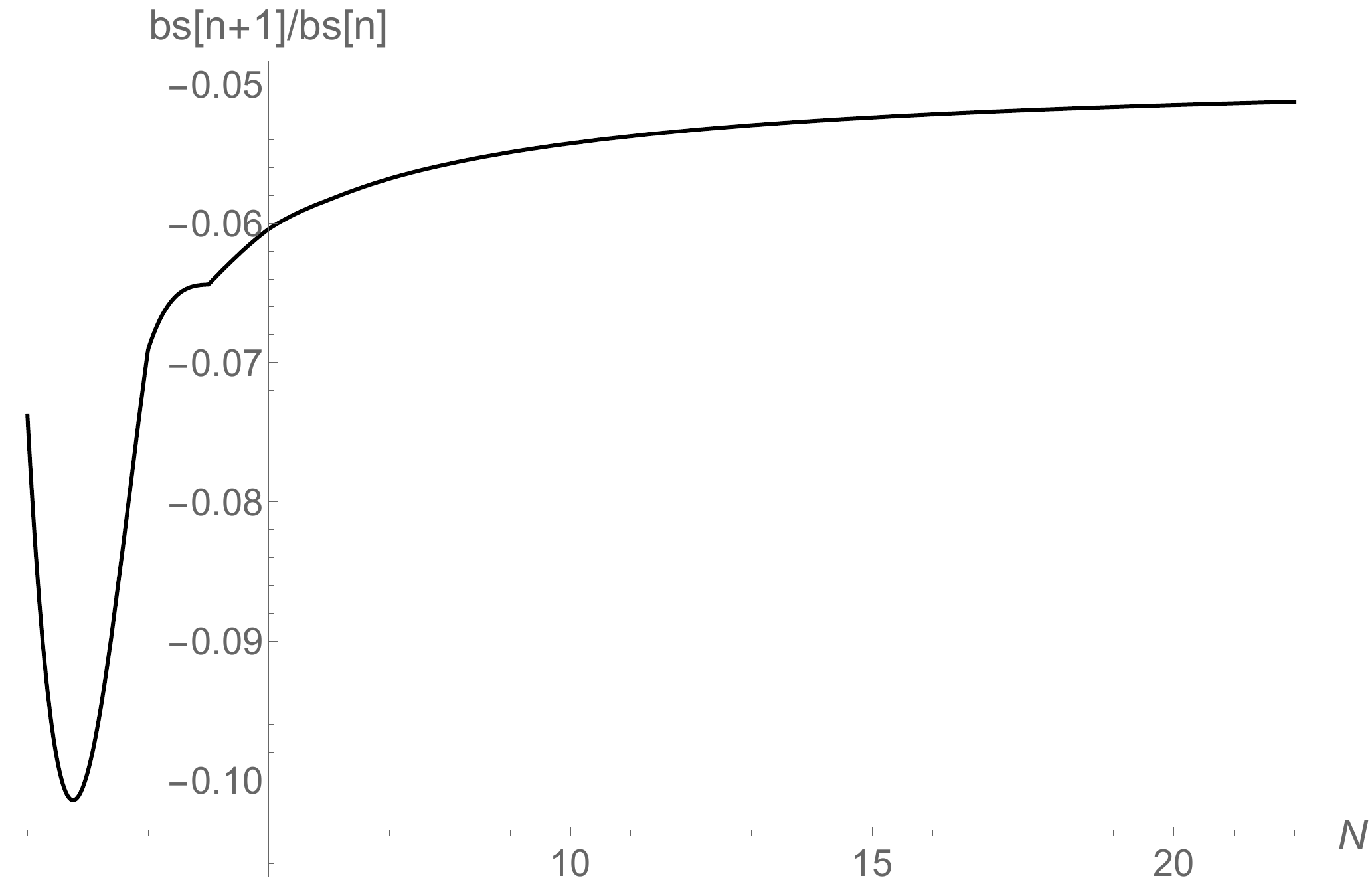}
\end{center}
\caption{The ratio of the coefficients $bs[n]$: $R=bs[n+1]/bs[n]$}
\label{PT}
\end{figure}

There are two main conclusions that one can make analyzing these plots. First of all,  that the PT works pretty well in the interval between zero and the first pole and the more terms are taken into account the better. In the case when we know the exact solution as for $\Sigma_A$ we can convince ourselves that the series is indeed convergent. In the case of $\Sigma_{sB}$ we check it numerically plotting the ratio of the coefficients $bs[n+1]/bs[n]$ (see Fig.\ref{PT}).

One can see that after some first orders it goes to the limit equal to $\approx -0.05$ which indicates the geometric progression type behaviour. Thus, the perturbation series seems to be convergent independently of the sign alternation.

The second conclusion concerns the singularity of the obtained functions. The total contribution of the leading and subleading  divergences for the ladder type diagrams in all orders can be written as
\beq
\Sigma_{Ladder}= \Sigma_A(z)+\epsilon (\Sigma_{sB}(z)+\frac{t}{s} \Sigma_{tB}(z))+ \dots, \ \ \ z\equiv \frac{g^2 s^2}{\epsilon}.
\eeq
From the analytical solution for $ \Sigma_A(z)$ (\ref{sol}) and $\Sigma_{tB}$ (\ref{sol2}) we see that while $\Sigma_{A}$
has an infinite sequence of poles, $\Sigma_{tB}$ exponentially grows although slower than $\Sigma_{A}$.
At the same time,  from the numerical solution for $\Sigma_{sB}$ it follows  that it apparently inherits the  singularities of
$\Sigma_{A}$ and does not cancel them. Thus, one can conclude that the subleading divergences do not change the
pattern of the leading ones.

\section{Discussion}

Summarizing the presented results concerning the subleading divergences one should stress once more that,
as it follows from the general theorems, the all loop terms are governed by the two loop subleading contributions.
We have presented explicit formulas confirming this statement for the ladder case and have demonstrated how the higher terms of PT can be calculated from the lower ones via pure algebraic recursive relations. The corresponding equations for the sum of all loop contributions (\ref{eq1},\ref{eq},\ref{Ric},\ref{Ric2}) generalize the usual
renormalization group relations for the pole terms for the case of non-renormalizable interactions.

To make everything explicit and transparent we chose the  set of the ladder type diagrams and performed the summation of the leading and subleading divergences in all loops. Even this task happened to be quite complicated
and we were bound to use numerical methods.  However, the result of summation of subleading divergences
does not lead to any qualitative difference from the leading terms. All the main features of the leading divergences keep untouched.

As for the total set of diagrams, already the leading divergences described by eq.(\ref{tot}) are difficult to analyze.
Following the same lines one can construct an analogous equation for the subleading divergences but it will be even more complicated.  The key question here is whether
the account of all diagrams will improve the behaviour of the laddrer type diagrams and remove the infinite sequence of poles or not.  More general question is to make sense of non-renormalizable interactions and to interpret the obtained results. Does the limit when $\epsilon \to 0$ exists or not? Has the infinite sequence of poles, if it survives, anything to do with the infinite number of bound states like in a string theory or not? \cite{stringtofield}
We leave the analysis of these questions to further publications.

\section*{Acknowledgements}
 The authors are grateful to M.Kompaniets for numerical check of the 3- and 4-loop ladder type diagrams. and to L.Bork for numerous useful discussions. Financial support from RFBR grants \# 14-02-00494 and \#16-32-00737 is acknowledged.

\end{document}